\documentclass[3p,times]{elsarticle}
\usepackage{ecrc}

\volume{00}

\firstpage{1}

\journalname{Annals of Physics}

\runauth{S. Longhi {\it et al.}}

\usepackage{amssymb}
\usepackage[figuresright]{rotating}

\begin{document}

\begin{frontmatter}



\title{Invisible defects in complex crystals}


\author{Stefano Longhi}
\address{Dipartimento di Fisica, Politecnico di Milano and Istituto di Fotonica e Nanotecnologie del Consiglio Nazionale delle Ricerche, Piazza L. da Vinci 32, I-20133 Milano, Italy\\ Tel/Fax: 0039 022399 6156/6126, email: longhi@fisi.polimi.it}
\author{Giuseppe Della Valle}
\address{Dipartimento di Fisica, Politecnico di Milano and Istituto di Fotonica e Nanotecnologie del Consiglio Nazionale delle Ricerche, Piazza L. da Vinci 32, I-20133 Milano, Italy}

\begin{abstract}
We show that invisible localized defects, i.e. defects that can not be detected by an outside observer, can be realized in a crystal with an engineered imaginary potential at the defect site. The invisible defects are synthesized by means of supersymmetric (Darboux) transformations of an ordinary crystal using band-edge wave functions to construct the superpotential. The complex crystal has an entire real-valued energy spectrum and Bragg scattering is not influenced by the defects. An example of complex crystal synthesis is presented for the Mathieu potential.
\end{abstract}


\end{frontmatter}

\section{Introduction}

Transport, scattering and trapping of matter or classical waves in periodic or disordered potentials is a fundamental issue in different areas of physics. As real-valued potentials are ubiquitous to describe Bragg scattering in ordinary crystals, in recent years an increasing attention has been devoted to 
study the transport properties of lattices described by an imaginary potential, that we refer here to as {\it complex crystals} (see, for instance, \cite{CC0,CC1,CC2,CC3,CC4,CC5,CC6,CC7,CC8,CC9}  and references therein).
Imaginary potentials have been introduced in
many physical contexts to describe the dynamics of reduced
systems \cite{Moiseyev,Rotter}, and found great interest in non-Hermitian
quantum mechanical frameworks, noticeably in systems described by $\mathcal{PT}$-symmetric Hamiltonians \cite{Bender}. 
In particular, wave scattering from an imaginary periodic potential has been investigated in a series of recent works for both matter \cite{CC2,CC3,CC4} and optical \cite{CC5,CC7,CC9} waves, highlighting some unusual and universal features as compared to ordinary crystals. Among others, we mention the violation of the Friedel's law of Bragg
scattering \cite{CC2,CC3}, double refraction and nonreciprocal diffraction \cite{CC5},
unidirectional Bloch oscillations \cite{CC8} and unidirectional invisibility \cite{CC9,Azana}.
In optics, synthetic complex crystals can be realized by 
introduction of a periodic index and balanced gain/loss modulations in a dielectric medium \cite{CC5,Azana,NP,Feng12}, whereas in atom optics experiments complex crystals have been realized  exploiting
the interaction of near resonant light with an open two-level
system \cite{CC3,CC4}. Recently, an elegant and experimentally accessible optical setting, based on temporal pulse propagation in coupled fiber loops, has been proposed \cite{Peschel} to mimic discretized dynamics of complex lattices, with the demonstration of most of the anomalous transport properties of complex crystals mentioned above.\\ 
The properties of complex crystals in the presence of defects or disorder are attracting a great attention as well \cite{dis1,dis2,dis3,Longhi,dis4,dis5,dis6}. A defect in an otherwise periodic potential is known to introduce a bound state, whose energy generally falls  inside an energy gap of the crystal. In most cases, a defect is not transparent to Bloch waves, causing partial reflection and partial transmission of a particle wave packet that propagates in the lattice. However, in some special cases it is known that the defect can be reflectionless.  The existence of reflectionless potentials for Hermitian systems was investigated in a pioneering work
by Kay and Moses in 1956 \cite{Kay}, and then studied in great
detail in the context of the inverse scattering theory
\cite{inv1,inv2} and supersymmetric quantum mechanics \cite{susy1}. The potentials obtained by such techniques,
though being transparent, are generally not {\it invisible}, i.e. they can be detected by an outside observer by e.g. time-of-flight measurements. This is due
to the dependence of the phase of the transmitted wave on energy,
which is generally responsible for some delay and/or for the
distortion of a wave packet transmitted across the potential
\cite{note1}. The experimental observation of reflectionless (but not invisible) defects in an Hermitian  tight-binding lattice model has been recently reported in Ref.\cite{Suk}. 
Reflectionless potentials, constructed by supersymmetric
 transformations, can be also realized in the non-Hermitian framework \cite{susyV}, however 
 the issue of invisibility of defects in complex crystals has not received great attention so far.
 In a recent work \cite{Longhi}, it has been theoretically suggested that invisible defects
can be synthesized in a {\it non-Hermitian} tight-binding lattice model, i.e. in a tight-binding lattice with complex-valued site energies and hopping rates.  
As tight-binding lattice models are appropriate to describe the transport properties of a narrow-band crystal, they do not account for the full band structure of the lattice and can not describe, for instance, multiband wave packet propagation and interband effects \cite{note2}. Hence, a natural question arises whether invisibility can be established, at least in principle, in a general model of complex crystal, i.e. beyond the single-band approximation. In this work we show that invisible defects can indeed be realized in a complex crystal, without recurring to any approximate model.  To this aim, we apply supersymmetric quantum mechanical  methods (Darboux transformation) to periodic potentials, which have been successfully employed to construct 
new solvable isospectral periodic potentials (see, for instance, \cite{PRD}) as well as $\mathcal{PT}$-symmetric potentials (see, e.g., \cite{susyannals} and references therein). To synthesize invisible defects,  the supersymmetric potential is built from a {\it complex} combination of periodic (Bloch) and unbounded  (i.e., not belonging to the spectrum) band-edge eigenfunctions of the lowest band of a reference Hermitian crystal. An example of crystal synthesis is presented for the Mathieu potential, taken as the reference Hermitian crystal.\\
The paper is organized as follows. In Sec.II we briefly review the main mathematical tools of supersymmetric quantum mechanics applied to periodic potentials, whereas in Sec.III  a very general procedure to synthesize complex crystals with invisible defects is presented. An example of defect invisibility, including direct numerical simulations of wave packet propagation in the synthesized lattice, is presented in Sec.IV. Finally, in Sec.V the main conclusions are outlined.  

\section{Supersymmetric periodic potentials}
In quantum physics,  the supersymmetric
quantum mechanics and other equivalent procedures, such as
the Darboux transformation or the factorization method, have been extensively used to construct exactly or quasi-exactly solvable
potentials of the Schr\"{o}dinger equation (see, for instance, \cite{susy1,darb} and references therein). In particular, supersymmetric quantum mechanical models with periodic potentials have been mainly investigated in the Hermitian case with the aim of constructing new solvable isospectral potentials (see, for instance, chapter 7 of \cite{susy1} and \cite{PRD}).  In this section we briefly review the supersymmetric technique for the Schr\"{o}dinger equation with a periodic potential as it has been commonly used \cite{PRD}, whereas an extension of the technique to synthesize complex crystals with invisible defects will be presented in the next section.\\
The motion of a particle wave packet in a one-dimensional periodic potential is governed by the Schr\"{o}dinger equation (with $\hbar=2m=1$)
\begin{equation}
 i \partial _t \psi(x,t)=\hat{H}_0 \psi(x,t)
 \end{equation}
  with the Hamiltonian
\begin{equation}
\hat{H}_0=-\frac{\partial^2}{\partial x^2}+V_0(x)
\end{equation}
 where $V_0(x)$ is the periodic potential with lattice period $a$ [$V_0(x+a)=V_0(x)$]. Since we are dealing with complex crystals, in our analysis we will quite generally allow the potentials to be complex valued.  Let us first remind a few properties of the spectrum of $\hat{H}_0$ for a real-valued potential $V_0$ \cite{periodic}.  In this case, the spectrum is purely continuous and composed by a set of energy bands, delimited by the energies $E_0$, $E_1$, $E_2$, $E_3$, ..., as shown in Fig.1. For a given value of the energy $E$ inside an allowed band, the Bloch theorem ensures the existence of two linearly-independent eigenfunctions of $\hat{H}_0$ belonging to the continuous spectrum, $\psi_{1,2}(x)=u_{1,2}(x) \exp( \pm i k x)$, where $u_{1,2}(x+a)= u_{1,2}(x)$ are periodic functions of $x$ with the same period $a$ of the lattice and $k$ in the Bloch particle quasi-momentum, which varies in the first Brillouin zone $ -\pi/a < k \leq  \pi/a$, with $k \neq 0, \pm \pi/a$. For an energy value $E$ corresponding to an edge of an allowed band, i.e. for $k=0, \pm \pi/a$ and $E=E_0, E_1, E_2, ...$, the two linearly-independent solutions to the Schr\"{o}dinger equation are given by \cite{periodic}
 \begin{equation}
 \psi_1(x)=u_1(x) \; , \; \psi_2(x)=u_2(x)+\frac{x}{a}u_1(x)
 \end{equation}
where $u_{1,2}(x+a)=u_{1,2}(x)$ for $k=0$ (i.e. for $E=E_0, E_3, E_4, ...$) and $u_{1,2}(x+a)=-u_{1,2}(x)$ for $k= \pm \pi/a$ (i.e. for $E=E_{1}, E_{2},E_5,  ...$). Of such two solutions, only the former one, $\psi_1(x)$, belongs to the continuous  spectrum of $\hat{H}_0$, whereas $\psi_{2}(x)$ is unlimited as $x \rightarrow \pm \infty$. Since $V_0(x)$ is real-valued, the functions $u_{1,2}(x)$ can be taken to be real-valued. Moreover, the oscillation theorem of the Hill equation \cite{Hill} ensures that, for  the lowest energy $E_0$, the periodic band-edge wave function $\psi_1(x)$ has no nodes. In the following analysis and without loss of generality, we will assume $E_0=0$.\par
\begin{figure}[b]
\includegraphics[width=8cm]{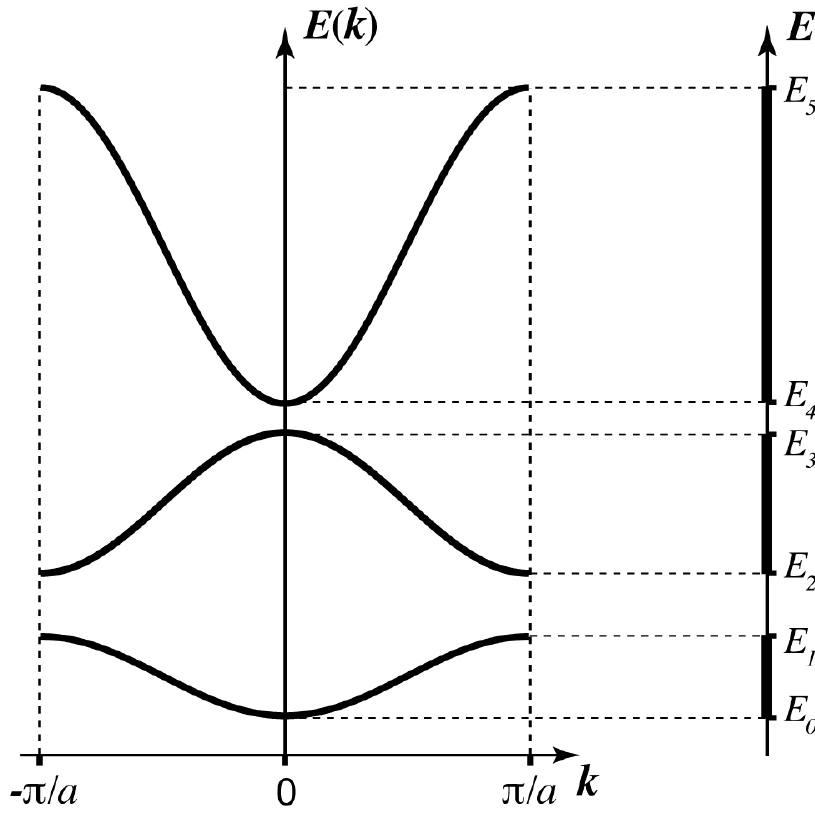}
\caption{(Color online). Schematic of the band structure of a one-dimensional lattice with  a periodic and real-valued potential $V(x)$. The band edge energies, starting from the lowest band edge, are denoted by $E_0$, $E_1$, $E_2$, $E_3$, ....  At such energies, two linearly-independent solutions $\psi_1(x)$ and $\psi_2(x)$ of the Schr\"{o}dinger equation have the form given by Eq.(3).}
\label{implementatio}
\end{figure}
Darboux transformations enable to  easily construct an Hamiltonian $\hat{H}_{1}$ with a potential $V_1(x)$, which is isospectral to $\hat{H}_0$. To this aim, the Hamiltonian $\hat{H}_0$ is factorized as follows
\begin{equation}
\hat{H}_0=\hat{B} \hat{A}
\end{equation}  
where 
\begin{equation}
\hat{A}= -\frac{\partial}{\partial x}+W(x) \; , \; \hat{B}=\frac{\partial}{\partial x}+W(x).
\end{equation}
In Eq.(5), $W(x)$ is the so-called superpotential, which satisfies the Riccati equation
\begin{equation}
V_0(x)=W^2(x)+\frac{dW}{dx}.
\end{equation}
The superpotential $W(x)$ can be determined from the relation
\begin{equation}
W(x)=\frac{1}{\phi(x)}\frac{d \phi}{d x}
\end{equation}
 where $\phi(x)$ is a solution to the equation 
 \begin{equation}
 -\frac{d^2 \phi}{d x^2}+V_0(x) \phi(x)=0,
 \end{equation}
 i.e. $\hat{H}_0 \phi=0$. The most general solution to Eq.(8) is given by an arbitrary superposition of the two linearly-independent band-edge functions $\psi_1(x)$ and $\psi_2(x)$, defined by Eqs.(3), for $E=E_0=0$. To avoid singularities of the superpotential $W(x)$, $\phi(x)$ has to be chosen such as $\phi(x) \neq 0$ on the entire real $x$ axis. We will come back later on this point. Let us then construct the partner Hamiltonian $\hat{H}_1=\hat{A} \hat{B}$, obtained from $\hat{H}_0$ by intertwining the operators $\hat{B}$ and $\hat{A}$.
 We note that, if the potential $V_0$ is real-valued, i.e. $\hat{H}_0$ is Hermitian, one has $\hat{B}=\hat{A}^{\dag}$ and the Darboux transformation  is equivalent to the intertwining method of supersymmetric quantum mechanics.\\
The following properties can be readily proven:\\
 (i) The potential $V_1(x)$ of the partner Hamiltonian $\hat{H}_1=-\partial^2_x+V_1(x)$ is given by
 \begin{equation}
 V_1(x)=W^2(x)-\frac{dW}{dx}=V_0(x)- 2 \frac{d^2}{dx^2} \left[ {\rm ln \;} \phi(x) \right].
 \end{equation}
(ii)  If $\psi(x)$ is an eigenfunction of $\hat{H}_0$ with eigenvalue $E \neq E_0=0$, then 
\begin{equation}
\xi(x)= \hat{A} \psi= \left[ -\frac{d}{dx}+W(x)  \right] \psi(x)
\end{equation}
is an eigenfunction of $\hat{H}_1$ with the same eigenvalue $E$. Moreover, Eq.(10) can be inverted, yielding
\begin{equation}
\psi(x)= \frac{1}{E} \left[ \frac{d}{dx}+W(x) \right] \xi(x). 
\end{equation}
(iii) For $E=E_0=0$, the two linearly-independent solutions to Eq.(8) with the potential $V_0(x)$ replaced by $V_1(x)$ are given by
\begin{equation}
f(x)=\frac{1}{\phi(x)} \; ,  \;  g(x)= \frac{1}{\phi(x)} \int_{0}^{x} d \xi \phi^2( \xi) 
\end{equation}
i.e. $\hat{H}_1 f=\hat{H}_1 g=0$.
Moreover
\begin{equation}
W(x)=-\frac{1}{f(x)} \frac{df}{dx}.
\end{equation}
As mentioned above, the function $\phi(x)$ that defines the superpotential $W(x)$ via Eq.(7) is not uniquely defined, and it is given by an arbitrary superposition of the band-edge functions $\psi_1(x)$ and $\psi_2(x)$, i.e.
\begin{equation}
\phi(x)  =\alpha \psi_1 (x)+ \beta \psi_2(x) 
 =  \left( \alpha + \beta \frac{x}{a} \right) u_1(x)+\beta u_2(x) 
\end{equation}
with $u_{1,2}(x+a)=u_{1,2}(x)$ and $u_{1}(x) \neq 0$ on the real axis $x$. If we confine ourselves to consider Hermitian Schr\"{o}dinger equations, to ensure a real-valued potential $V_{1}(x)$ of the partner Hamiltonian $\hat{H}_1$ the constants $\alpha$ and $\beta$ entering in Eq.(14) must be real-valued. Moreover, to avoid singularities in the  potential $V_1(x)$ the function $\phi(x)$ should not vanish on the real axis, which necessarily implies $\beta=0$. Hence, apart from the inessential multiplication factor  $\alpha$, the function $\phi(x)=u_1(x)$ is periodic with the same period $a$ as $V_0(x)$. The potential $V_1(x)$ is thus periodic and given by [see Eq.(9)]
\begin{equation}
V_1(x)= 2 \left[ \frac{1}{u_1(x)} \frac{du_1}{dx} \right]^2-\frac{1}{u_1(x)} \frac{d^2u_1}{dx^2}
\end{equation}
Moreover, since $f(x)=1 / \phi(x)$ is a periodic function and $\hat{H}_1 f=0$, $E=E_0=0$ belongs to the continuous spectrum of $\hat{H}_1$, i.e. the two periodic crystals defined by $\hat{H}_1$  and $\hat{H}_2$ are {\it isospectral}. Such a property of supersymmetric periodic potentials has been discussed, for example, in Refs. \cite{susy1,PRD} and used to construct new exactly solvable periodic potentials.
\begin{figure}[b]
\includegraphics[width=8cm]{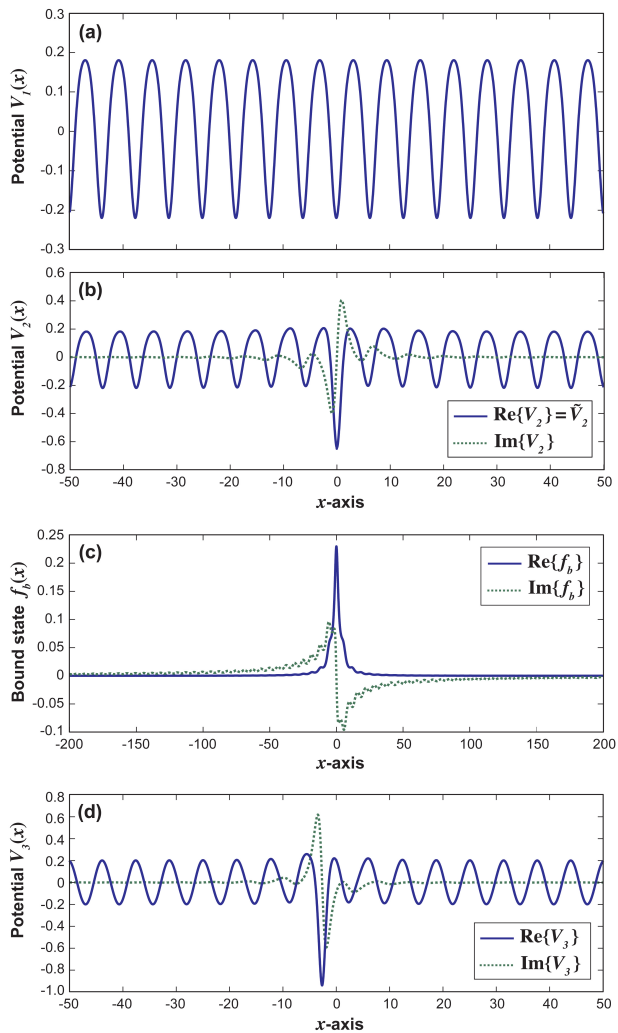}
\caption{(Color online). Synthesis of a complex lattice with an invisible defect by Darboux transformations applied to the Mathieu potential (30), taken  as a reference potential $V_0(x)$. Parameter values are $V_0=0.2$ and $a=2 \pi$. (a) Behavior of the periodic and real-valued potential $V_1(x)$, defined  Eq.(15). (b) Behavior of the complex-valued lattice $V_2(x)$ (real and imaginary parts) with a defect near $x=0$, as given by Eq.(18) for $\gamma=2$. The bound state at energy $E=0$ sustained by the defect, as given by Eq.(19), is shown in panel (c). (d) Behavior of the complex-valued lattice $V_3(x)$ (real and imaginary parts) with a defect near $x=0$, as given by Eq.(27) for ${\rm Re}(\lambda)={\rm Im}(\lambda)=10 \times {\rm max} \; \{ u_1(x) \}$.}
\end{figure}

\section{Synthesis of complex crystals with invisible defects}
To synthesize periodic potentials with invisible defects, in this section we extend the supersymmetric method briefly reviewed in the previous section by allowing the supersymmetric potential (and hence the partner potential) to be complex-valued.
 We proceed into two steps.\par
 In the first step, let us assume a reference real-valued periodic potential $V_0(x)$, and let us construct the isospectral real-valued partner potential $V_1(x)$, given by Eq.(15) and obtained by assuming the superpotential $W(x)=(1 / \phi) (d \phi/ dx)$ with $\phi(x)=u_1(x)$, as discussed in the last part of the previous section.\par
  In the second step,  let us exploit the non-uniqueness   of the superpotential when the coefficients $\alpha$ and $\beta$ in Eq.(14) are allowed to be complex-valued, and let as assume the following superpotential
  \begin{equation}
  \tilde{W}(x)=\frac{1}{\tilde{\phi}(x)} \frac{d \tilde{\phi}}{d x }
  \end{equation}
  with 
  \begin{equation}
  \tilde{\phi}(x)  =  \psi_1(x)+i \gamma \psi_2(x) \\
   =  \left( 1+i \gamma \frac{x}{a}  \right)u_1(x)+i \gamma u_2(x) 
  \end{equation}
  where $\gamma$ is an arbitrary real-valued parameter.  The partner potential $V_2(x)$ generated by the superpotential $\tilde{W}(x)$ is complex-valued and reads explicitly
  \begin{equation}
  V_2(x)  =  2 \left[ \frac{1}{\tilde{\phi}(x)} \frac{d \tilde{\phi}}{dx} \right]^2-\frac{1}{\tilde{\phi}(x)} \frac{d^2 \tilde{\phi}}{dx^2} 
   =  2 \left[  \frac{(i \gamma/a)u_1+(1+i \gamma x/a) u^{'}_1+i \gamma u_{2}^{'}}{(1+i \gamma x /a) u_1+i \gamma u_2)} \right]^2 
  -  \frac{2 i (\gamma/a)u_{1}^{'} +(1+i \gamma x /a) u_{1}^{''}+i \gamma u_{2}^{''}} {(1+i \gamma x/a)u_1+i \gamma u_2} 
  \end{equation}
  where $u_{1}^{'}= (du_1/dx)$, $u_{1}^{''}= (d^2 u_1/dx^2)$, etc. Note that, for $\gamma=0$ the potential $V_2(x)$ reduced to $V_1(x)$ given by Eq.(15). For $\gamma \neq 0$, the potential $V_2(x)$ is complex valued and {\it not} periodic. However, one can readily show that, far from $x=0$, namely at $x \rightarrow \pm \infty$, one has $V_{2}(x) \rightarrow V_1(x)$. This means that $V_2(x)$ reproduces the periodic potential $V_1(x)$ but with a {\it defect} near $x=0$. Near $x \sim 0$, the potential $V_2(x)$ has a non-vanishing imaginary part, i.e. $V_2(x)$ describes a crystal with a {\it complex defect}. For construction, the Hamiltonian $\hat{H}_2=-d^2 /dx^2+V_2(x)$ of the complex crystal with the defect is  isospectral to $\hat{H}_0$, and thus to $\hat{H}_1$, apart from the fact that the energy $E=E_0=0$ belongs to the {\it point} spectrum of $\hat{H}_2$, rather than to the continuous spectrum as for $\hat{H}_1$ and $\hat{H}_0$. Indeed, according to property (iii) mentioned in Sec.II,  the function
 \begin{equation}
  f_{b}(x)=\frac{1}{ \tilde{\phi}(x)}=\frac{1}{\left( 1+i \gamma x/a \right)u_1(x)+i \gamma u_2(x) }
 \end{equation}
  is a normalizable solution  to the equation $\hat{H}_2 f_{b}(x)=0$, and hence $E=0$ belongs to the point spectrum of $\hat{H}_2$. In other words, the defect introduces a {\it bound state} at the bottom $E=E_0=0$ of the crystal energy spectrum.  One can then readily prove that the defect is invisible to any Bloch wave packet that propagates along the lattice, i.e. it does neither reflect the wave packet nor introduces any distortion or delay. In fact, let us consider an energy $E \neq 0$ taken inside an allowed energy band of the crystal, and let us indicate by $f_0(x)=u_0(x) \exp(i kx)$ the Bloch eigenfunction, with wave number $k$, of the crystal with potential $V_0(x)$, i.e. $\hat{H}_0 f_0(x)=E f_0(x)$. According to Eqs.(11) and (12), the corresponding eigenfunctions $f_1(x)$ and $f_2(x)$ of the crystals with potentials $V_1(x)$ and $V_2(x)$ are given by
 \begin{eqnarray}
 f_0(x) & = & \frac{1}{E} \left(  \frac{d}{dx}+\frac{u_{1}^{'}}{u_1} \right) f_1(x) \\
 f_2(x) & = & \left( -\frac{d}{dx}+\frac{\tilde{\phi}^{'}}{\tilde{\phi}} \right) f_0(x),
 \end{eqnarray} 
  where $\tilde{\phi}(x)$ is given by Eq.(17).  Substitution of Eq.(20) into Eq.(21) yields
 \begin{equation}
f_2(x)  =  \frac{1}{E} \left( -\frac{d}{dx}+\frac{\tilde{\phi}^{'}}{\tilde{\phi}} \right) \left(  \frac{d}{dx}+\frac{u_{1}^{'}}{u_1} \right) f_1(x).
 \end{equation} 
 Note that, since $V_1(x)$ is a periodic potential, $f_1(x)$ is of Bloch-Floquet type, whereas $f_2(x)$ {\it is not} of Bloch-Floquet type because the potential $V_2(x)$ is not periodic. 
 However, in view of the asymptotic behavior of $\tilde{\phi}(x)$ and $\tilde{\phi}^{'}(x)$ as $x \rightarrow \pm \infty$,  i.e. $\tilde{\phi}(x) \sim i \gamma (x/a) u_1(x) $ and  $\tilde{\phi}^{'}(x) \sim i \gamma (x/a) u_{1}^{'}(x) $ as $ x \rightarrow \pm \infty$, it readily follows that
  \begin{equation}
  f_2(x)  \sim  \frac{1}{E} \left[- \frac{d^2}{dx^2}- \left( \frac{u_{1}^{'}}{u_1} \right)^{'}  + \left( \frac{u_{1}^{'}}{u_1}\right)^2 \right] f_1(x) 
   =  \frac{1}{E} \left[- \frac{d^2}{dx^2}+V_1(x) \right] f_1(x)=f_1(x)  
  \end{equation}
  i.e. 
  \begin{equation}
  f_2(x) \sim f_1(x) \;\;\;\; {\rm as} \;\;\;\; x \rightarrow  \pm \infty
  \end{equation}
  Any wave packet propagating in the two crystals $V_1(x)$ and $V_2(x)$  can be decomposed as a superposition (integral) of 
  the Bloch-Floquet eigenfunctions $f_1(x)$ and $f_2(x)$, belonging to the various lattice bands. Hence  Eq. (24) ensures that any forward-propagating wave packet in the complex lattice $V_2(x)$, which at initial time is localized far from the defect on the left hand side [where $f_1(x) \simeq f_2(x)$], after crossing the defect spreads like in the periodic (defect-free) Hermitian lattice $V_1(x)$, i.e. the defect in $V_2(x)$ is fully invisible. We remark that such a result holds for a wave packet with energy components belonging to {\it any} band of the lattice, i.e. defect invisibility is not restricted to a single band. \par
  As a final comment, it should be noted that the procedure outlined above allows one to synthesize an invisible defect embedded in the periodic potential $V_1(x)$, which is a susy partner of the original (reference) periodic potential $V_0(x)$. A slightly modified procedure can be used to synthesize an invisible defect embedded into the original periodic potential $V_0(x)$, i.e. a complex potential $V_3(x)$ which is isospectral to $V_0(x)$ and differs from it for an invisible defect localized at around $x=0$. To synthesize the potential $V_3(x)$, let us notice that , according to Eqs.(12), two different superpotentials $W_1(x)$ and $\tilde{W}_1(x)$ associated to $V_1(x)$ are given by $W_1(x)=(1/f) (df/dx)=- (1/ \phi ) (d \phi /dx)=-W(x)$ and $\tilde{W}_1(x)=(1 / \chi) (d \chi /dx)$, where $\chi(x)=\lambda f(x) +g(x)$, $\lambda$ is an arbitrary complex number, and the functions $f(x)$, $g(x)$ are defined by Eq.(12).  Using the superpotentials $W_1(x)$ and $\tilde{W}_1(x)$, one can then construct two Hamiltonians which are isospectral to $\hat{H}_1=-\partial^2_x+V_1(x)$. Since $W_1(x)=-W(x)$, the former Hamiltonian is precisely $\hat{H}_0$ with the periodic potential $V_0(x)$. The latter Hamiltonian, $\hat{H}_3=-\partial_x^2+V_3(x)$, is associated to the potential 
  \begin{equation}
  V_3(x)=\tilde{W}_1^2(x)-\frac{d \tilde{W}_1}{dx}=V_1(x)-2 \frac{d^2}{dx^2} \left[ { \rm{ln}} \; \chi(x) \right].
  \end{equation}
  Using Eqs.(9) and Eq.(12), from Eq.(25) one obtains
  \begin{equation}
  V_3(x)=V_0(x)-2 \frac{d^2}{dx^2} \left[  {\rm{ln}} \; \chi(x) \phi(x)  \right]= V_0(x)-2 \frac{d^2}{dx^2} \left[  {\rm{ln}} \; \left( \lambda+\int_0^x d \xi \phi^2( \xi)  \right)  \right].
  \end{equation}
  Assuming $\phi(x)=u_1(x)$, one then obtains after straightforward calculations the following expression for $V_3(x)$
  \begin{equation}
  V_3(x)=V_0(x)-\frac{4 u_1(x) (du_1 /dx)}{\lambda+ \int_0^x d \xi u_1^2( \xi)} +\frac{2u_1^4(x)}{\left[ \lambda+\int_0^x d \xi u_1^2( \xi)  \right]^2}.
  \end{equation}
  Note that $V_3(x) \rightarrow V_0(x)$ as $x \rightarrow \pm \infty$, and that $V_3(x)$ is not singular for any {\it complex} number $\lambda$. Moreover, it can be readily shown that, for construction,  the defect embedded in the potential $V_3(x)$ is invisible and sustains a bound state, given by 
  \begin{equation}
 f_b(x) =\frac{u_1(x)}{\lambda+ \int_0^x d \xi u_1^2( \xi)}
  \end{equation}
    with energy $E=E_0=0$. The potential $V_3(x)$ thus provides a one-parameter complex potential that differs from the original potential $V_0(x)$ for an invisible defect at around $x=0$.

  \section{An example of complex crystal with an invisible defect}
  As an example of crystal synthesis, let us consider as a reference potential the sinusoidal potential (Mathieu potential)
  \begin{equation}
  V(x)=V_0 \cos (2 \pi x /a),
  \end{equation}
  which is a well-studied and soluble problem in energy band theory \cite{Sla}. 
The band structure and corresponding Bloch functions of the Mathieu potential can be readily determined making use of the Fourier expansion of the wave function (the momentum eigenfunction) and using the method of continued fractions (see, for instance, \cite{Sla,Parra}).  For example, for $a=2 \pi$ and $V_0=0.2$, the band edge energies $E_0$, $E_1$, $E_2$, ... (see Fig.1) are found to be given by $E_0=-0.0197$, $E_1=0.1452$, $E_2=0.3447$,  $E_3=0.9967$, $E_4=1.0163$, ...   Let us then assume as a reference potential $V_0(x)$ the shifted potential
\begin{equation}
V_0(x)=V_0 \cos (2 \pi x/a)-E_0
\end{equation}
and let us indicate by $u_1(x)$ and $u_2(x)$ the two periodic functions defined by Eqs.(3) and related to the two linearly-independent solutions $\psi_1(x)$, $\psi_2(x)$  to the Mathieu equation $-(d^2 \psi/dx^2) +[V_0 \cos (2 \pi x /a)-E_0] \psi=0$. Owing to the inversion symmetry $V_0(-x)=V_0(x)$ of the potential $V_0(x)$, the two functions $u_1(x)$ and $u_2(x)$ have a well-defined parity (even and odd for $u_1$ and $u_2$, respectively). With the functions $u_1(x)$ and $u_2(x)$, we can then construct the two lattices with potentials $V_1(x)$ and $V_2(x)$, defined by Eqs.(15) and (18), respectively. 
\begin{figure}[b]
\includegraphics[width=8cm]{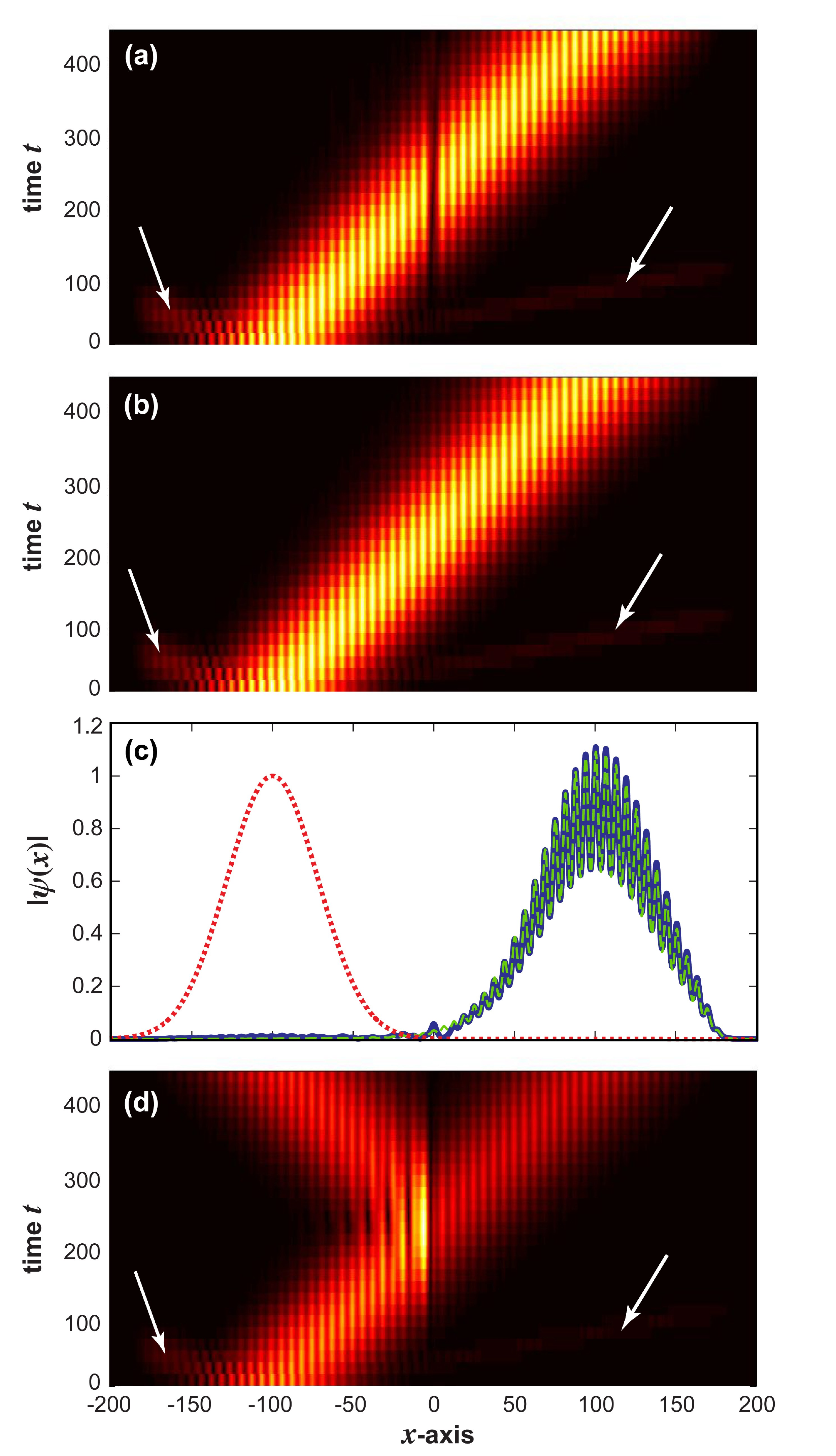}
\caption{(Color online). (a) Wave packet evolution in the complex lattice $V_2(x)$ of Fig.2(b) with an invisible defect (snapshot of $|\psi(x,t)|$) for the initial condition defined by Eq.(27) and for parameter values $x_0=-100$, $k_0=1/4$ and $w=40$. The initial wave packet mostly excited the lowest band of the lattice, though small excitation of higher-order bands is visible (the weaker wave packets highlighted by the arrows in the figure). (b) Same as (a), but for the real-valued periodic lattice $V_1(x)$ of Fig.2(a). (c) Behavior of $|\psi(x,t)|$ at time $t=450$ corresponding to the cases (a) [solid curve] and (b) [dashed curve, almost overlapped with the solid one]. The dotted curve centered at $x=x_0=-100$ depicts the initial wave packet profile $| \psi(x,0)|$. 
(d) Same as (a), but for the real-valued lattice potential $V_3(x)={\rm Re} \{ V_2(x) \}$ with a non-invisible defect.}
\end{figure}
Figures 2(a) and 2(b) show the profiles of the two potentials for  $a=2 \pi$, $V_0=0.2$ and $\gamma=2$. For construction, $V_1(x)$ is a real-valued and periodic potential, whereas $V_2(x)$ is the associated complex-valued periodic potential which differs from $V_1(x)$ owing to the defect at around $x=0$, see Fig.2(b). We note that, owing the inversion symmetry of the functions $u_1(x)$ and $u_2(x)$, the complex potential $V_2(x)$ turns out to be $\mathcal{PT}$ symmetric \cite{noteq}.  In Fig.2(c) we show the bound state $f_b(x)$ at energy $E=0$, sustained by the potential $V_2(x)$ and given by Eq.(19). We have checked the invisibility of the defect in the complex lattice $V_2(x)$ by comparison of the propagation of a wave packet in the two lattices $V_1(x)$ and $V_2(x)$. To this aim, we numerically solved the Schr\"{o}dinger equations $i \partial_{t} \psi(x,t)=-\partial^{2}_{x} \psi(x,t)+V_{1,2}(x) \psi(x,t) $ using an accurate pseudospectral method for parameter values $V_0=0.2$, $a=2 \pi$ and $\gamma=2$. As an initial condition, we assumed a Gaussian wave packet, of width $w$ and localized at the position $x_0<0$ far from the defect, with a mean momentum $k_0$, i.e. we assumed
\begin{equation}
\psi(x,0)=\exp[-(x-x_0)^2/w^2] \exp(ik_0x).
\end{equation}
\begin{figure}[b]
\includegraphics[width=8cm]{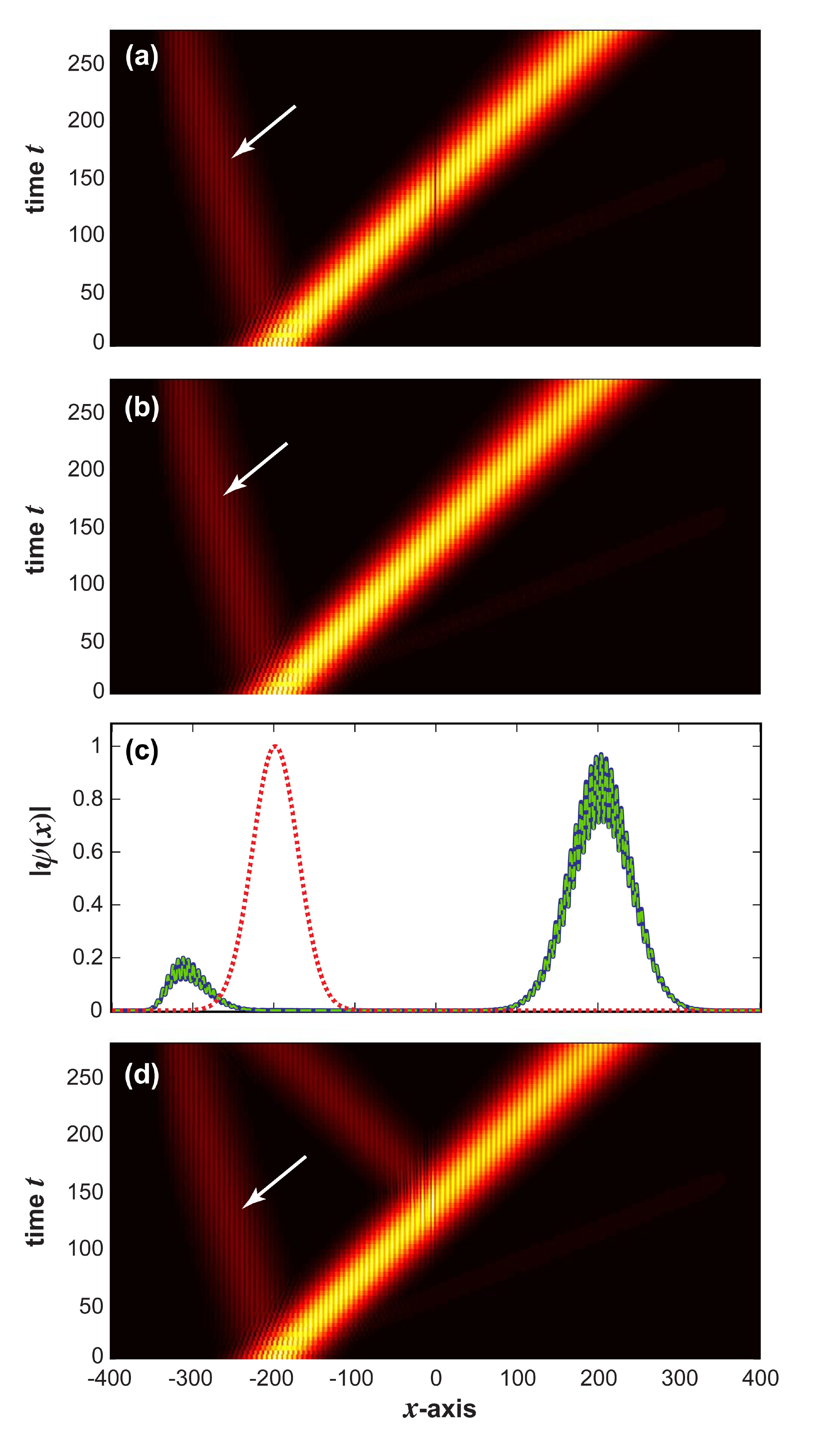}
\caption{(Color online). Same as Fig.3, but for  parameter values $x_0=-200$, $k_0=3/4$ and $w=40$. In this case the initial wave packet mainly excites the second  band of the lattice, though small excitation of the lowest band is visible  (the weaker wave packet highlighted by the arrows in the figures).}
\end{figure}
The input wave packet (31) generally excites Bloch-Floquet states belonging to different bands of the lattice, with a weight that is determined by the so-called Bloch excitation functions $B_l(k_0)$, where $l$ is the lattice band index (see, for instance, \cite{kepalle}). The group velocities and group velocity dispersion of the wave packet components belonging to the various lattice bands are determined by the derivatives of the band dispersion curves at $k=k_0$. Hence,  the initial wave packet (27) generally breaks into several wave packets that propagate at different speeds and undergo different spreading \cite{kepalle}.  For $|k_0|< \pi/a=1/2$, the lowest lattice band is mostly excited, and a forward propagating wave packet requires additionally $k_0>0$. An example of wave packet propagation in the two lattices $V_1(x)$ and $V_2(x)$  corresponding to excitation the  lowest lattice band is shown in Figs.3(a) and (b) for parameter values $x_0=-100$, $w=40$ and $k_0=1/4$.  A comparison of Figs.3(a) and (b) clearly shows that the defect in the complex lattice $V_2(x)$ is invisible, i.e. the wave packet in the lattice $V_2(x)$ crosses the defect without being reflected nor being delayed or distorted as compared to the corresponding wave packet propagating in the defect-free lattice $V_1(x)$. This is clearly shown in Fig.3(c), which depicts the shape of $|\psi(x,t)|$ at the time $t=450$. A different defect would  rather generally partially reflect the incident wave packet and distort the transmitted one. For example, in Fig.3(c) we show the numerically-computed wave packet evolution in the potential $\tilde{V}_2(x)={\rm Re}\{V_2(x) \}$, obtained from $V_2(x)$ by neglecting the imaginary part of the potential.  In this case  the wave packet is partially scattered off by the defect.\\ 
As predicted by the theoretical analysis presented in the previous section, invisibility of the defect for the potential $V_2(x)$ is not restricted to the excitation of the lowest lattice band. This is shown, as an example, in Fig.4 where the numerically-computed wave packet propagation is depicted in the three lattices $V_1(x)$, $V_2(x)$ and $\tilde{V}_2(x)$ for parameter values  $x_0=-200$, $w=40$ and $k_0=3/4$. In this case, the initial distribution $\psi(x,0)$ excites mainly the second lattice band, and the corresponding wave packet propagates forward. A wave packet component belonging to the lowest band is visible as well, which now propagates backward. The plots of Fig.4(c) clearly show that the defect is invisible for a wave packet belonging to the second lattice band as well. \par
As a final comment, we note that in our example we have discussed the invisibility of the complex potential $V_2(x)$, which basically reproduces the periodic potential $V_1(x)$ with an embedded defect near $x=0$. As shown in the last part of the previous section, one could in a similar way synthesize a complex potential $V_3(x)$, defined by Eq.(27), which coincides with the original Mathieu potential $V_0(x)=V_0 \cos (2 \pi x /a)-E_0$ but with an invisible defect. An example of the potential $V_3(x)$, as defined by Eq.(3), is shown in Fig.2(d) for parameter values $V_0=0.2$, $a=2 \pi$ and ${\rm Re}(\lambda)={\rm Im}(\lambda)=10 ;\ {\rm max} \; \{ u_1(x) \}$. Note that, as compared to $V_2(x)$, the potential $V_3(x)$ is not $\mathcal{PT}$ invariant.

\section{Conclusion}
 In this paper we have theoretically shown the existence of invisible defects in complex crystals, i.e. lattices described by a complex-valued potential. Such crystals have been synthesized using Darboux transformations, following a similar procedure used to realize reflectionless defects  in Hermitian Hamiltonians.  However, as reflectionless defects in Hermitian systems  are not invisible, the defects synthesized in a complex potential can be invisible, i.e. their existence can not be detected by an outside observer. The complex crystal with an invisible defect is obtained by a double Darboux transformation applied to a  reference Hermitian crystal and using as a superpotential a {\it complex} combination of periodic (Bloch) and unbounded (not belonging to the Hamiltonian spectrum) band-edge eigenfunctions of the lowest band. The synthesis procedure ensures that the defect is invisible to any Bloch wave packet propagating in the lattice with energy components belonging to the various lattice bands. An example of lattice synthesis and wave packet propagation has been presented for the Mathieu potential. It is envisaged that our results could be extended to synthesize different kinds of invisible defects in complex lattices, for example by cascading  successive Darboux transformations. Also, our idea could be extended to relativistic electrons using Darboux transformations for the Dirac (rather than the Schr\"{o}dinger) wave equation.
 Finally, our study could suggest further directions of research,  for example the realization of reflectionless or even invisible interfaces between two isospectral (but different) crystals, which might be of relevance to the physics and properties of interfaces.


\begin{thebibliography}{00}
\bibitem{note2}
A well-known and important example where a tight-binding lattice model fails to properly describe  coherent wave transport in a lattice is provided by 
 Bloch oscillations, which raised a long-standing controversy about the existence of the Wannier-Stark ladder.
\bibitem{noteq}
It should be noted that, with the technique of the Darboux transformation, the potential $V_2(x)$ does not necessarily turns out to be $\mathcal{PT}$-invariant, as in the example discussed in Sec.IV. For example, assuming $V_0(x)=V_0 \cos (2 \pi x /a)+V_0 \sin( 6 \pi x/a)-E_0$ (rather than the Mathieu potential) as a reference potential, the resulting complex potential $V_2(x)$, given by Eq.(18),  is not $\mathcal{PT}$-invariant, in spite its spectrum is real-valued for construction.
\end{thebibliography}
\end{document}